\documentclass[article,9pt]{IEEEtran}
\usepackage{amsmath}
\usepackage{amsfonts}
\usepackage{amssymb}
\usepackage{setspace}
\usepackage{graphicx}
\usepackage{amsthm}
\usepackage{color}
\usepackage{cite}
\usepackage{bm}
\usepackage{epsfig,psfrag}
\usepackage{tabularx}
\usepackage{tikz}
\usepackage{pst-node}
\usepackage{acronym}
\usepackage[yyyymmdd,hhmmss]{datetime}
\usepackage{notation}
\usepackage{transparent}
\usepackage{subcaption}
\usetikzlibrary{arrows,backgrounds,calc,positioning,shapes,shadows}

\usepackage{tikzscale}

\allowdisplaybreaks
\DeclareMathOperator*{\argmin}{arg\,min}

\newcolumntype{L}[1]{>{\raggedright\arraybackslash}p{#1}}
\newcolumntype{C}[1]{>{\centering\arraybackslash}p{#1}}
\newcolumntype{R}[1]{>{\raggedleft\arraybackslash}p{#1}}

\providecommand{\ist}{\hspace*{.3mm}}

\providecommand{\rmv}{\hspace*{-.3mm}}
\providecommand{\iist}{\hspace*{1mm}}

\DeclareMathAlphabet{\mathpzc}{OT1}{pzc}{m}{it}

\acrodef{pnt}[PNT]{positioning, navigation and timing}
\acrodef{pa}[PA]{physical anchor}
\acrodef{va}[VA]{virtual anchor}
\acrodef{mva}[MVA]{master virtual anchor}
\acrodef{slam}[SLAM]{simultaneous localization and mapping}
\acrodef{crb}[CRB]{Cram\'{e}r-Rao bound}
\acrodef{crlb}[CRLB]{Cram\'{e}r-Rao lower bound}
\acrodef{speb}[SPEB]{squared position error bound}
\acrodef{fim}[FIM]{Fisher information matrixe}
\acrodef{mi}[MI]{mutual information}
\acrodef{rdm}[RDM]{ranging direction matrix}
\acrodef{rdv}[RDV]{ranging direction vector}
\acrodef{is}[IS]{information-seeking}
\acrodef{pmf}[pmf]{probability mass function}
\acrodef{pdf}[pdf]{probability density function}
\acrodef{mmse}[MMSE]{minimum mean-square error}
\acrodef{mse}[MSE]{mean-square error}
\acrodef{rh}[RH]{receding horizon}
\acrodef{ekf}[EKF]{extended Kalman filter}
\acrodef{ukf}[UKF]{unscented Kalman filter}
\acrodef{roi}[ROI]{region of interest}
\acrodef{bp}[BP]{belief propagation}

\newdateformat{monthyeardate}{\monthname[\THEMONTH] \THEDAY, \THEYEAR} 

\newcommand{\paperTitle}{Active Planning for Cooperative Localization: \\
A Fisher Information Approach
}
\newcommand{\paperTitleMarkboth}{}

\begin{document}
\title{\paperTitle \vspace*{3mm}}

\author{\normalsize Wenyu~Zhang\IEEEauthorrefmark{1},~\IEEEmembership{\normalsize Student Member,~IEEE} and Bryan~Teague\IEEEauthorrefmark{2},\hspace{-.3mm} \IEEEmembership{\normalsize Member,\hspace{-.3mm} IEEE},
Florian~Meyer\IEEEauthorrefmark{1},\hspace{-.3mm} \IEEEmembership{\normalsize Member,\hspace{-.3mm} IEEE}\\[3mm]
\IEEEauthorblockA{\IEEEauthorrefmark{1}University of California San Diego, La Jolla, CA, Email: \{wez078, flmeyer\}@ucsd.edu \vspace*{0mm}}\\[.3mm]
\IEEEauthorblockA{\IEEEauthorrefmark{2}MIT Lincoln Laboratory, Lexington, MA, USA, Email: bryan.teague@ll.mit.edu \vspace*{-3.5mm}}}

\maketitle

\begin{abstract}
    Location-aware networks will introduce new services and applications for modern convenience, surveillance, and public safety.  In this paper, we consider the problem of cooperative localization in a wireless network where the position of certain anchor nodes can be controlled. We introduce an active planning method that aims at moving the anchors such that the information gain of future measurements is maximized. In the control layer of the proposed method, control inputs are calculated by minimizing the traces of approximate inverse Bayesian \acp{fim}. The estimation layer computes estimates of the agent states and provides Gaussian representations of marginal posteriors of agent positions to the control layer for approximate Bayesian FIM computations. Based on a cost function that accumulates Bayesian \ac{fim} contributions over a sliding window of discrete future timesteps, a \ac{rh} control is performed. Approximations that make it possible to solve the resulting tree-search problem efficiently are also discussed. A numerical case study demonstrates the intelligent behavior of a single controlled anchor in a 3-D scenario and the resulting significantly improved localization accuracy.
\end{abstract}


\acresetall
\section{Introduction}
\label{sec:introduction}

Location awareness in wireless networks is an important aspect in a variety of applications in search and rescue, agriculture monitoring, and maritime surveillance. In cooperative localization, the nodes of a wireless network communicate and perform distance measurements to estimate their position and possibly further motion-related parameters. The localization accuracy in the network strongly depends on the quality of the wireless links between nodes and the network topology. Thus, in many applications, it is advantageous to control properties of the network such as the positions of certain nodes or the communication parameters of the wireless links \cite{Bul:09,Cor:10}.
The problem of combining estimation and control in wireless networks is often referred to as ``controlled networked sensing'' \cite{Bul:09,Cor:10}. In this work, we focus on node position control for cooperative localization. In particular, we aim to develop a method that can actively plan the motion of certain well-localized mobile nodes such that the information gain of future measurements is maximized. 

\subsection{State of the Art}
In controlled networked sensing, a control policy is used to maximize the information gained from observations with respect to the unknown states of interest. Possible measures of information gain include negative posterior entropy and \ac{mi} \cite{HofCla:10, Jul:12, Ata:14, MeyWymHla:15,MeyWymFroHla:J15} as well as scalar-valued functions of the \ac{fim} \cite{Mor:13}. Methods that rely on negative posterior entropy or \ac{mi} are suitable for nonlinear and non-Gaussian system models but perform computationally intensive operations based on particle representations of probability distributions. In case the unknown states are high-dimensional, due to the curse of dimensionality, these particle-based operations are unfeasible. \ac{fim}-based methods, on the other hand, are more computationally efficient but rely on Gaussian representations of probability distributions that are typically less accurate \cite{Mor:13,MeyWymFroHla:J15}. 

\begin{figure}
\centering
\includegraphics[scale=.48]{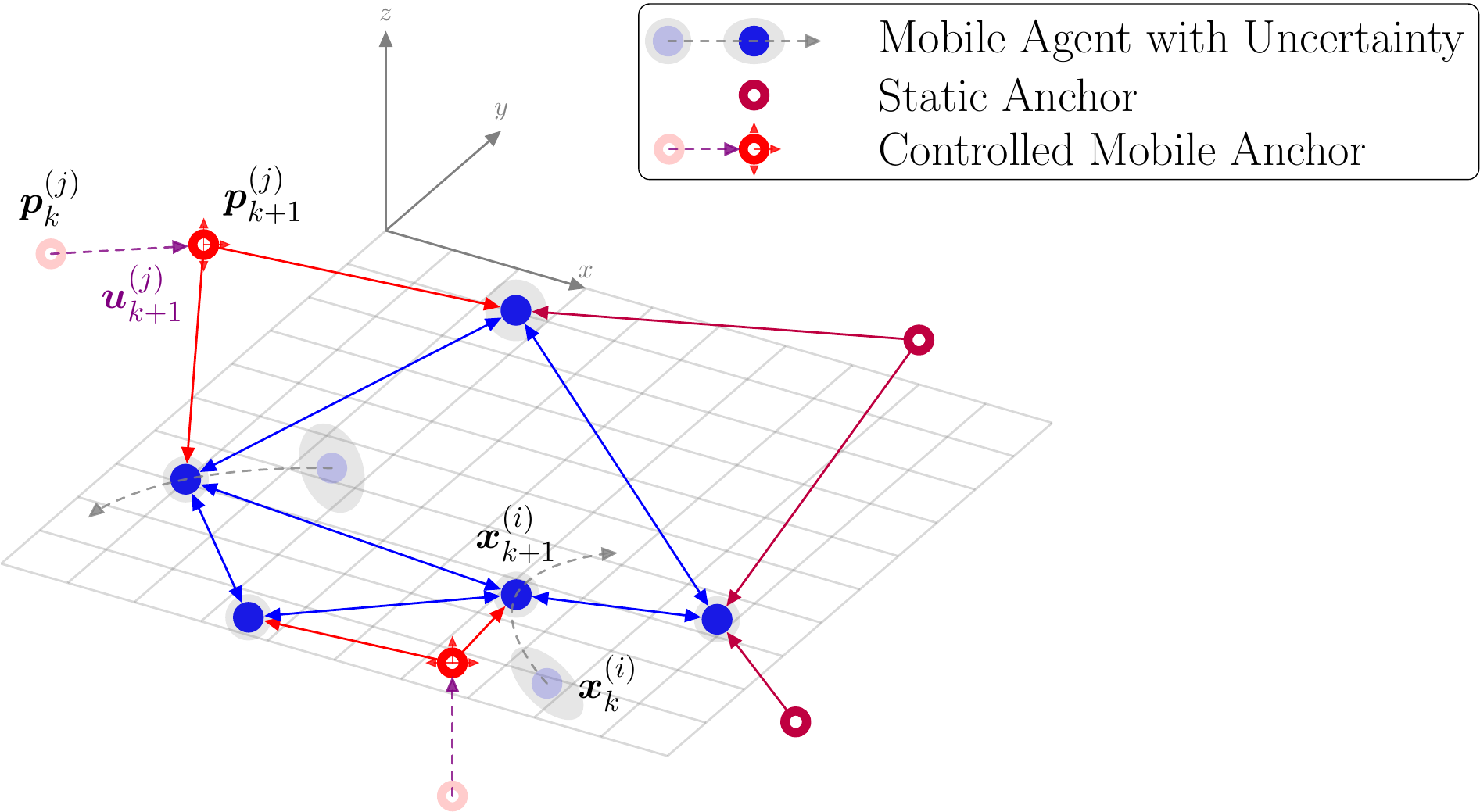}
\centering
\caption{\small Cooperative localization scenario with mobile agents, mobile anchors, and static anchors.}\label{fig:coopL}
\vspace{0mm}
\end{figure}
Due to the distributed nature of the mobile agent network, conventional sequential Bayesian estimation methods \cite{HofCla:10, Jul:12, Ata:14, MeyWymHla:15,MeyWymFroHla:J15, Mor:13} are unsuitable for state estimation in cooperative localization scenarios. State-of-the-art methods in this setting rely on loopy \ac{bp} which enables distributed and scalable sequential Bayesian estimation \cite{WymLieWin:J09,MeyWymFroHla:J15}. While most existing work on estimation methods for cooperative localization assume that nodes in the network are static, a variety of emerging networks, e.g., in the context of the Internet-of-Things\cite{WinMeyLiuDaiBarCon:J18}, search-and-rescue\cite{Tea:22}, and underwater surveillance\cite{FerMunTesBraMeyPelPetAlvStrLeP:J17}, consist of mobile nodes. This calls for the development of controlled networked sensing approaches for cooperative localization.

In \cite{MeyWymHla:15}, BP-based cooperative localization is combined with myopic information-seeking control of some nodes in the network. Here, information is quantified by the negative conditional differential entropy \cite{Cov:06} of all unknown states conditioned on all measurements collected by the network. For each controlled node, a myopic control policy based on partial derivatives of the negative conditional differential entropy is developed. The computation of partial derivatives is performed using Monte Carlo integration and is thus very costly. In addition, the resulting gradient-based myopic control policy is unsuitable for \ac{rh} control \cite{MeyWymFroHla:J15}. To the best of our knowledge, active planning, i.e., information-seeking control over a receding planning horizon, is still an open research problem in a cooperative localization\vspace{-1mm} setting.

\subsection{Contributions, Paper Organization, and Notation}
In this paper, we introduce active planning for cooperative localization networks that aims at increasing the information gain of future measurements. Contrary to \cite{MeyWymHla:15}, which develops a control policy based on negative conditional differential entropy, we propose a control policy that minimizes traces of approximate inverse Bayesian FIMs \cite{SheWin:10}.  In this way, costly Monte Carlo integration can be avoided, and the computational complexity related to a myopic control, i.e., a control strategy with a planning horizon of one timestep, can be strongly reduced.

Due to this reduced computational complexity, we can develop a \ac{rh} control, i.e., a control strategy with a planning horizon that consists of multiple future timesteps. In particular, we develop a cost function that accumulates Bayesian FIM contributions and state the tree search problem that results from cost function minimization. Approximations for solving the resulting tree-search problem efficiently are also discussed. The key contributions of this work are as\vspace{1.5mm} follows.
\begin{itemize}
\item We introduce the trace of the inverse Bayesian \ac{fim} as the cost function for active planning in cooperative\vspace{1.5mm} localization.
\item We develop a cost function that accumulates Bayesian FIM contributions for \ac{rh}\vspace{1.5mm} control.
\item We demonstrate that the proposed method can improve localization performance in a numerical case\vspace{1.5mm} study.
\end{itemize}
Note that in this preliminary work, we use an approximation of the Bayesian \ac{fim} that only takes the measurements of the other anchors into account. Developing an approximation of the Bayesian \ac{fim} that also takes the measurements of the other agents into account \cite{SheWymWin:10} is subject to future work.

\section{System Model and Estimation Problem}
We consider a wireless network for cooperative localization in 3-D. The network consists of agents with unknown positions and anchors with known positions.  An example network is shown in Fig. \ref{fig:coopL}. The agents and some anchors are mobile. The neighborhood of an agent is defined by its communication range. Mobile agents communicate with other mobile agents and anchors in their neighborhood to exchange position information and perform pairwise measurements of the distance. The goal of each mobile agent is to sequentially estimate its location by using all available pairwise distance measurements and the position information of anchors and mobile agents in its neighborhood. While the motion of agents is arbitrary, it is assumed that the motion of the mobile anchors can be controlled and planned to improve localization accuracy in the network. In what follows, we introduce the system model, formulate the joint estimation and control problem, and review the estimation layer of the proposed estimation and control method.

\subsection{System Model}
\label{sec:systemModel}
We consider a wireless network that consists of $M$ anchors and $N$ mobile agents. There are $M_1$ mobile anchors indexed $j \rmv\in \{1,\dots,M_1\}$ and $M_2 = M\rmv-\rmv M_1$ static anchors indexed by $j \in \{M_1\rmv+\rmv1,\dots,M\}$. The state of agent $i \rmv\in\rmv \{1,\dots,N\}$ at discrete timestep $k$ is denoted as $\V{x}_{k}^{(i)}$ and consists of the agent's 3-D position as well as further motion-related parameters. The 3-D position of anchor $j \rmv\in\rmv \{1,\dots,M\}$ at discrete timestep $k$ is denoted as $\V{p}_{k}^{(j)} \rmv=\rmv \big[ p_{1,k}^{(j)} \ist\ist\ist p_{2,k}^{(j)} \ist\ist\ist p_{3,k}^{(j)} \big]^{\rmv \mathrm{T}}\rmv\rmv$.

Agent motion is described by a potentially nonlinear state-transition model, i.e.,
\begin{equation}
    \label{eq:statetransitionf}
    \V{x}_{k+1}^{(i)} = f(\V{x}_{k}^{(i)}) + \nu_{k}^{(i)},\quad i \in \{1,\dots,N\}
\end{equation}
where $f(\V{x}_{k}^{(i)})$ is the state transition function and $\nu_{k}^{(i)}$ is zero-mean Gaussian driving noise with standard deviation $\sigma_{\nu}$.

Anchor motion is described by the velocity control vector $\V{u}^{(j)}_{k+1} \in \mathcal{U}$, i.e.\vspace{-1mm},
\begin{equation}
    \label{eq:controlf}
    \V{p}_{k+1}^{(j)} = \V{p}_{k}^{(j)} + T_{\rmv\Delta} \ist \V{u}^{(j)}_{k+1},\quad j \in \{1,\dots,M_1\}
    \vspace{1mm}
\end{equation}
where $T_{\rmv\Delta}$ is the duration of a discrete timestep. Note that $\mathcal{U}$ is a dictionary that consists of a finite number of possible control vectors. The position of static anchors remains fixed, i.e., $\V{p}_{k+1}^{(j)} = \V{p}_{k}^{(j)}\rmv\rmv$, $j \rmv\in\rmv \{M_1 \rmv+\rmv 1, \dots, M\}$, $\forall k$. 

The set of all anchors that are in the neighborhood of agents $i$ is denoted as $\Set{M}_i \subset \{1,\dots,M\}$. Agent $i$ performs pairwise distance measurements with all anchors $j \rmv\in\rmv \Set{M}_i$\vspace{1mm}, i.e.,
\begin{equation}
    \label{eq:measurementf1}
    \V{z}_{k}^{(i,j)} = \big\|  \mathpzc{x}_{\hspace{.4mm} k}^{(i)}- \V{p}_{k}^{(j)} \big\| + \omega^{(i,j)}_k, \quad j \in \mathcal{M}_i
    \vspace{1mm}
\end{equation}
where $\mathpzc{x}_{\hspace{.4mm} k}^{(i)}  = \big[ x_{1,k}^{(i)}  \ist\ist\ist\ist x_{2,k}^{(i)} \ist\ist\ist\ist x_{3,k}^{(i)} \big]^{\mathrm{T}}$ denotes the position of agent $i$, and $\omega^{(i,j)}_k$ is zero-mean Gaussian measurement noise with standard deviation $\sigma^{(i,j)}_{\omega}\rmv\rmv$. We also introduce the set of all agents that are in the neighborhood of anchor $j \in \{1,\dots,M\}$, i.e., $\Set{A}_j = \{ \ist i \ist | \ist j \rmv\in\rmv \Set{M}_i \}$.

The set of all agents that are in the neighborhood of agents $i$ is denoted as $\Set{N}_i \subset \{1,\dots,N\} \backslash \{i\}$. Agent $i$ performs pairwise distance measurements with all agents $i' \rmv\in\rmv \Set{N}_i$\vspace{1mm}, i.e.,
\begin{align}
    \label{eq:measurementf2}
    \V{z}_{k}^{(i,i')} = \big\|  \mathpzc{x}_{\hspace{.4mm} k}^{(i)}  - \mathpzc{x}_{\hspace{.4mm} k}^{(i')}  \big\| + \eta^{(i,i')}_k, \quad i' \in \mathcal{N}_{i}
\end{align}
where $\eta^{(i,i')}_k$ is zero-mean Gaussian measurement noise with standard deviation $\sigma^{(i,j)}_{\eta}\rmv\rmv$. Typically, neighborhood sets can be assumed symmetric, i.e., $i' \rmv\in\rmv \mathcal{N}_{i} \iff i \in \mathcal{N}_{i'}$. At time $k=0$, it is assumed that prior \acp{pdf} of all agent states $f(\V{x}^{(i)}_0)$, $i \in \{1,\dots,N\}$ are known.

\begin{figure}[t!]
\centering
\includegraphics[scale=.205]{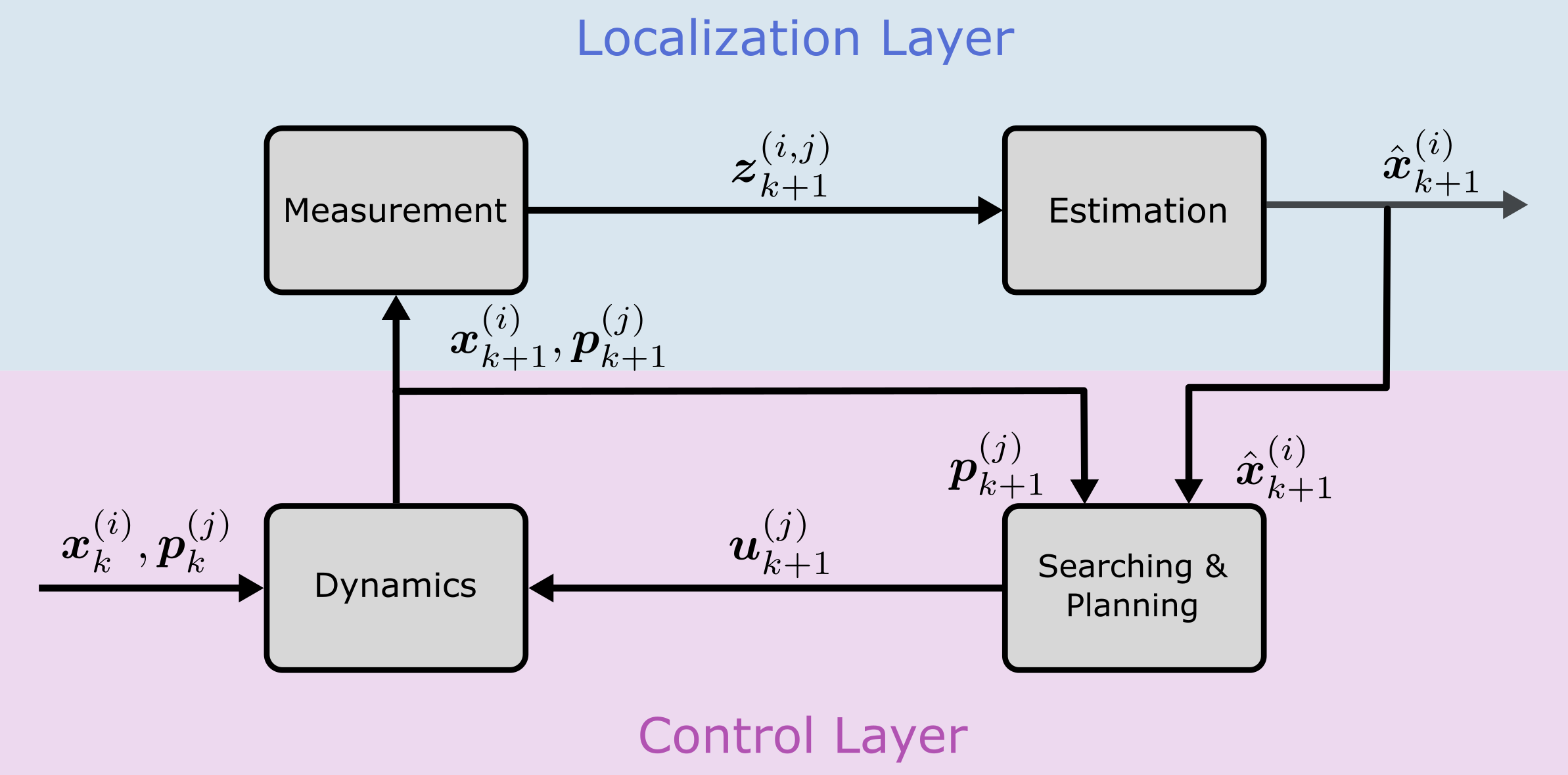}
\caption{\small Block diagram of the proposed estimation and control framework for cooperative\vspace{0mm} localization.}\label{fig.diagram}
\end{figure}

\subsection{Estimation for Cooperative Localization}
Fig. \ref{fig.diagram} shows a block diagram of the proposed estimation and control framework for cooperative localization. The ``dynamics'' block represents the state transition model of the agents in \eqref{eq:statetransitionf} and the control update model of the mobile anchors in \eqref{eq:controlf}. The ``measurement'' block represents the measurement models in \eqref{eq:measurementf1} and \eqref{eq:measurementf2}.

The ``measurement'' and ``estimation'' blocks constitute the localization layer. The main goal of the localization layer is to estimate the state of each agent based on all measurements collect up to the current time $k$. Let $\V{x}_{k} = \big[\V{x}_{k}^{(1) \mathrm{T}} \cdots \ist\ist \V{x}_{k}^{(N) \mathrm{T} }\big]^{\mathrm{T}}$ be the joint vector that consists of all agent states at time $k$ and let $\V{z}_{k}$ be the joint vector that consists of all measurements at time $k$, as defined in \eqref{eq:measurementf1} and \eqref{eq:measurementf2}. From the statistical model discussed in Section \ref{sec:systemModel}, the joint posterior \ac{pdf} $f(\V{x}_{k} | \V{z}_{1:k};\V{u}_{1:k})$ can be derived (see \cite{WymLieWin:J09} and \cite{MeyWymFroHla:J15} for details). Note that here we do not explicitly denote the known anchor positions.

State estimation of the agent $i$ can be performed by calculating the \ac{mmse}\vspace{.5mm} estimate \cite{Poo:B94} according to
\begin{equation}
    \hat{\V{x}}_{k}^{(i)}
    \ist\triangleq \int \V{x}_{k}^{(i)}\ist f(\V{x}^{(i)}_{k} | \V{z}_{1:k};\V{u}_{1:k}) \ist\mathrm{d}\V{x}_{k}
    \label{eq:mmseEst}
\end{equation}
where $f(\V{x}^{(i)}_{k} | \V{z}_{1:k};\V{u}_{1:k})$ is the marginal posterior \ac{pdf} of agent state $i$. Direct computation of this marginal posterior \ac{pdf} from the joint posterior \ac{pdf} $f(\V{x}_{k} | \V{z}_{1:k};\V{u}_{1:k})$ by marginalization is typically unfeasible. However, using common assumptions \cite{WymLieWin:J09,MeyWymFroHla:J15}, by exploiting the factorization structure of $f(\V{x}_{k} | \V{z}_{1:k};\V{u}_{1:k})$, it is possible to compute an accurate approximation of $f(\V{x}^{(i)}_{k} | \V{z}_{1:k};\V{u}_{1:k})$ in a fully distributed way. In particular, for distributed estimation of $\hat{\V{x}}_{k}^{(i)}\rmv\rmv$, agent $i$ only needs measurements that involve its state as well as position information from agents and anchors in its neighbor sets \cite{WymLieWin:J09,MeyWymFroHla:J15}.

Distributed estimation is typically performed using loopy \ac{bp} \cite{WymLieWin:J09,MeyWymFroHla:J15}. There is a wide range of \ac{bp} variants, which can be roughly classified into methods that represent marginal posterior \acp{pdf} by Gaussians and methods that represent marginal posterior \acp{pdf} by particles. Particle representations are typically more accurate compared to Gaussian representations but suffer from the curse of dimensionality \cite{DauHua:03}. Using a particle representation in the considered 3-D cooperative localization scenario would lead to an unfeasible computational complexity. For this reason, we make use of sigma point \ac{bp} \cite{MeyHliHla:J14}, which computes a Gaussian representation of marginal posterior \acp{pdf} based on the unscented transformation \cite{JulUhl:04}.

\section{Active Planning}
\label{sec:activePlanning}
In this section, we will introduce the main contribution of this paper, namely, active planning based on the \ac{fim}. The active planning method is represented by the ``searching \rmv \& \rmv planning'' block in Fig.~\ref{fig.diagram}. For active planning, a control policy is designed to move anchors such that future measurements increase the trace of the \ac{fim} also know as the \ac{crb}. In particular, the trace of the \ac{fim} for the considered problem is reviewed and then used to develop an accumulated cost function for \ac{rh} control. Finally, a near-optimal control vector is computed for each anchor by performing a tree search.

\begin{figure}[t]
\centering
\includegraphics[scale=.7]{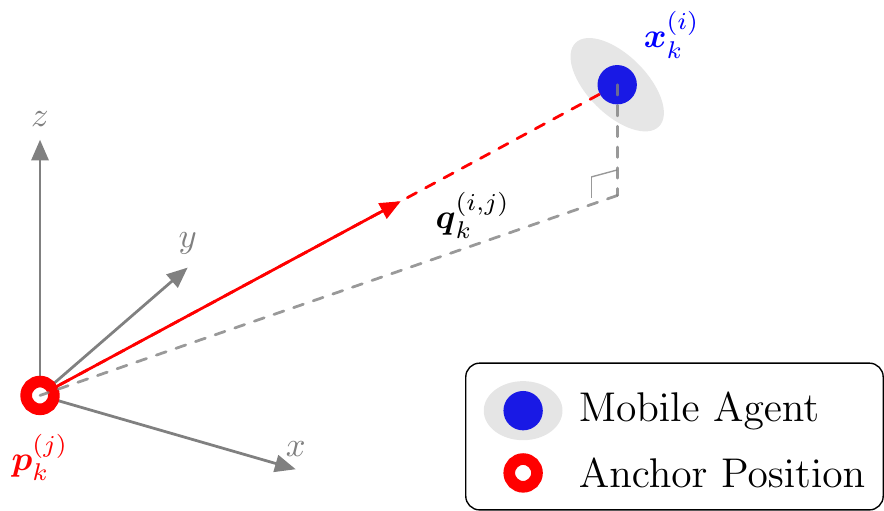}
\vspace{1.5mm}
\caption{\small 3-D range direction vector between position of anchor $j$ and position of agent $i$\vspace{0mm}.}
\label{fig:RD}
\end{figure}

\subsection{Equivalent \ac{fim} and Myopic Control Policy}
At time $k$, for each mobile anchor $j = 1,\dots,M_1$, we want to determine the control vector $\V{u}^{(j)}_{k+1}$ in \eqref{eq:controlf}, that maximizes the overall information provided by the anchor measurements performed at the next timestep, i.e., $\V{z}_{k+1}^{(i,j)}, i \in \mathcal{A}_j$. As a measure of information, we use the equivalent Bayesian \ac{fim} as derived in \cite{SheWin:10} that will be reviewed next.

Let us consider the range vector $\V{\mathpzc{q}}_k^{(i,j)} \! =\! \big [x_{1,k}^{(j)} - p_{1,k}^{(j)} \iist\iist x_{2,k}^{(j)} - p_{2,k}^{(j)} \iist\iist x_{3,k}^{(j)} - p_{3,k}^{(j)} \big]^\text{T}$ and the corresponding range direction vector $\V{q}_k^{(i,j)} \! =\! \V{\mathpzc{q}}_k^{(i,j)} \! / \| \V{\mathpzc{q}}_k^{(i,j)} \|$ from the position $\V{p}_{k}^{(j)}$ of anchor $j$ to the position $\mathpzc{x}_{\hspace{.4mm} k}^{(i)}$ of agent $i$, as shown in Fig.~\ref{fig:RD}. The \ac{rdm} \cite{SheWin:10} of agent $i$ with anchor $j$ can now be obtained as\vspace{.5mm}
\begin{equation}
    \V{J}\big(\mathpzc{x}_{\hspace{.4mm} k}^{(i)}\rmv\rmv, \V{p}_{k}^{(j)} \big) \triangleq \V{q}_k^{(i,j)} \big( \V{q}_k^{(i,j)} \big)^{\rmv \mathrm{T}} \rmv\rmv.
\end{equation}
Based on the \acp{rdm} of agent $i$ with all anchors $j \rmv\in\rmv \Set{M}_i$, an equivalent \ac{fim} \cite{SheWin:10} for the position of agent $i$ is given \vspace{.5mm} by
\begin{equation}
    \V{I}^{(i)} \rmv \big(\mathpzc{x}_{\hspace{.4mm} k}^{(i)}\rmv\rmv, \V{p}_{k} \big) = \sum_{j \in \Set{M}_i}\ist\frac{1}{\big(\sigma^{(i,j)}_{\omega}\big)^2} \ist\ist\ist \V{J}\rmv\big(\mathpzc{x}_{\hspace{.4mm} k}^{(i)}\rmv\rmv, \V{p}_{k}^{(j)}\big)
    \vspace{1mm}
\end{equation}
where we recall that $\sigma^{(i,j)}_{\omega}$ is the measurement standard deviation of the measurement noise $\omega^{(i,j)}_{k}$ in \eqref{eq:measurementf1}. 

Let us assume that a Gaussian representation of the predicted posterior \ac{pdf} for the agent state position $\mathpzc{x}_{\hspace{.4mm} k+1}^{(i)}$ is available, i.e.,  $f(\mathpzc{x}_{\hspace{.4mm} k+1}^{(i)} | \V{z}_{1:k};\V{u}_{1:k}) \rmv=\rmv  \mathcal{N}(\V{\mu}_{k+1}^{(i)},\V{\Sigma}_{k+1}^{(i)})$. This Gaussian representation can be provided by sigma point \ac{bp} \cite{MeyHliHla:J14} performed in the estimation layer. Now, we can obtain an approximate Bayesian \ac{fim} \cite{SheWin:10} based on the predicted mean $\V{\mu}_{k+1}^{(i)}$ and covariance $\V{\Sigma}_{k+1}^{(i)}$ of the agent position $\mathpzc{x}_{\hspace{.4mm} k+1}^{(i)}$,\vspace{.5mm} as
\begin{equation}\label{eq:fimGaussian}
    \V{I}_{k}^{(i)} \big( \V{p}_{k+1} \big)  = \V{I}^{(i)} \rmv \big(\V{\mu}_{k+1}^{(i)}, \V{p}_{k+1} \big) + \V{\Sigma}_{k+1}^{(i)^{-1}}\rmv.
    \vspace{1mm}
\end{equation}
Note that in this preliminary work, we use an approximation of the Bayesian \ac{fim} in \eqref{eq:fimGaussian}, which does not take the measurements among agents into account.

Next, let $\V{u}_{k+1} = \big[\V{u}_{k+1}^{(1)} \cdots \ist \V{u}_{k+1}^{(M_1)}\big]^{\text T}$ be\vspace{.2mm} the joint control vector and let $\V{\mathpzc{u}}_{k+1} = [\V{u}^{\text T}_{k} \ist\ist\ist \V{0}^{\text T}_{3 M_2}]^{\text T}$ be the augmented joint control vector. By using \eqref{eq:controlf} in \eqref{eq:fimGaussian}, we can now rewrite the approximate Bayesian \ac{fim} as a function as of $\V{u}_{k+1}$\vspace{1mm}, i.e.,
\begin{equation}\label{eq:fimGaussianControl}
    \V{C}_{k}^{(i)} \rmv (\V{u}_{k+1}) = \V{I}_{k}^{(i)} \big(  \V{p}_{k} \rmv+\rmv T_{\rmv\Delta}  \ist \V{\mathpzc{u}}_{k+1} \big).
    \vspace{1mm}
    \end{equation}
Here, we make use of the fact that at time $k$, all anchor positions $\V{p}_k$ are known and only the first $M_1$ anchors are mobile, i.e., the anchor topology only depends on the joint control vector $\V{u}_{k+1}$. 

To obtain a cost function for all agents states, we compute the sum of all traces of the individual approximate Bayesian \acp{fim},\vspace{.8mm} i.e.,
\begin{equation}\label{eq:costFunction}
    C_k\big(\V{u}_{k+1}\big) = \sum_{i=1}^{N} \text{Tr} \ist \Big(\V{C}_{k}^{(i)} \rmv (\V{u}_{k+1})^{-1} \Big).
    \vspace{.8mm}
\end{equation}
This sum is an approximation of the Bayesian \ac{crb} of the joint agent position $\mathpzc{x}_{\hspace{.4mm} k+1} \rmv\rmv=\rmv\rmv \big[\mathpzc{x}_{\hspace{.4mm} k+1}^{(1) \mathrm{T}} \cdots \ist\ist \mathpzc{x}_{\hspace{.5mm} k+1}^{(N) \mathrm{T} }\big]^{\mathrm{T}}$\rmv\rmv\rmv. The myopic control policy finally\vspace{.5mm} reads
\begin{equation}\label{eq:policy1}
    \hat{\V{u}}_{k+1}=  \argmin_{{\V{u}}_{k+1}\in \mathcal{U}^{\rmv M_1}}C_k\big(\V{u}_{k+1}\big).
    \vspace{-1mm}
\end{equation}



\subsection{\ac{rh} Control Policy and Tree-Search}\label{sec:RH}

Due to uncertain agent states, a myopic control policy as introduced in \eqref{eq:policy1}, can lead to highly suboptimal control decisions that also negatively influence estimation and control at future timesteps. This ``getting stuck'' at a local minimum often happens at initial timesteps where the localization uncertainty of the network is significant. To avoid local minima, we make use of \ac{rh} control \cite{Mat:11} which uses model-based prediction of future costs within a sliding window that consists of $T$ timesteps.

For \ac{rh} control, at the current time $k$, we aim to develop an accumulated cost function that considers $t\!\in\! \{1,\dots, T\}$ future timesteps. First, for all $T$ future timesteps and all agents, Gaussian representations for the predicted posterior \acp{pdf} of the agent state positions $\mathpzc{x}_{\hspace{.4mm} k+t}^{(i)}$ are obtained, i.e.,  we calculate $f(\mathpzc{x}_{\hspace{.4mm} k+t}^{(i)} | \V{z}_{1:k};\V{u}_{1:k}) \rmv=\rmv  \mathcal{N}(\V{\mu}_{k,t}^{(i)},\V{\Sigma}_{k,t}^{(i)})$, $t\!\in\! \{1,\dots, T\}$, $i\rmv\in\rmv\{1,\dots,N\}$. These Gaussian representations can be computed by the prediction step of a Kalman filter or a sigma point filter \cite{JulUhl:04}. Note that $\V{\mu}_{k,1}^{(i)}$ and $\V{\Sigma}_{k,1}^{(i)}$ are equal to $\V{\mu}_{k+1}^{(i)}$ and $\V{\Sigma}_{k+1}^{(i)}$ in \eqref{eq:fimGaussian}.  Next, based on \eqref{eq:fimGaussian}, we introduce the approximate Bayesian \ac{fim} of agent state $i$ for future timestep $k+t$, predicted at current timestep $k$,\vspace{-1.5mm} i.e.,

\begin{equation}\label{eq:fimGaussian1}
    \V{I}_{k,t}^{(i)} \big( \V{p}_{k,t} \big)  = \V{I}^{(i)} \rmv \big(\V{\mu}_{k,t}^{(i)}, \V{p}_{k,t} \big) + \V{\Sigma}_{k,t}^{(i)^{-1}}
    \vspace{.5mm}
\end{equation}
where $\V{p}_{k,t}$ is given by $\V{p}_{k,t} = \V{p}_{k} \rmv+\rmv T_{\rmv\Delta}  \sum^{t}_{t' = 1} \rmv \V{\mathpzc{u}}_{k+ t'}$. We can now rewrite the predicted approximate Bayesian \ac{fim} in \eqref{eq:fimGaussian1} as a function of $\V{u}_{k+1:k+t}$\vspace{1mm}, i.e.,
\begin{equation}\label{eq:fimGaussianControl1}
    \V{C}_{k,t}^{(i)} \rmv (\V{u}_{k+1:k+t}) = \V{I}_{k,t}^{(i)}  \bigg( \V{p}_{k} + T_{\rmv\Delta} \rmv \sum^{t}_{t' = 1} \rmv \V{\mathpzc{u}}_{k+ t'} \rmv \bigg).
    \vspace{.5mm}
 \end{equation}

To obtain a cost function that consists for all agents states, as in \eqref{eq:costFunction}, we  again compute the sum of all traces of the individual predicted approximate Bayesian \acp{fim},\vspace{.5mm} i.e.,
\begin{equation}\label{eq:costFunction1}
    C_{k,t}(\V{u}_{k+1:k+t}) = \sum_{i=1}^{N} \text{Tr} \ist \Big(\V{C}_{k,t}^{(i)} \rmv (\V{u}_{k+1:k+t})^{-1} \Big).
    \vspace{.5mm}
\end{equation}
Note that for $t = 1$, \eqref{eq:costFunction1} is equal to \eqref{eq:costFunction}.

To simplify the notation, we now drop the time index $k$. In particular, we write $\V{u}_{1:t}$ for $\V{u}_{k+1:k+t}$ and $C_{t}\big(\V{u}_{t};\V{u}_{1:t-1})$ for $C_{k,t}\big(\V{u}_{k+1:k+t})$ The accumulated cost function for a horizon $T\geq1$ based on the control sequence $\V{u}_{1},\dots,\V{u}_{T}$, is now expressed\vspace{.5mm} as
\begin{equation}\label{eq:accumulatedCost}
    C(\V{u}_{1:T}) = \sum_{t=1}^{T}\hspace{.3mm} \gamma^{t-1} C_{t}\big(\V{u}_{t};\V{u}_{1:t-1})
    \vspace{.5mm}
\end{equation}
where $0 \rmv<\rmv\gamma\rmv\leq\rmv1$ is the discount factor \cite{Put:B94}. The discount factor is motivated by the fact that at timesteps that are far in the future, predicted agent states are less certain, and thus the cost contributions related to these timesteps should have lower weights in the accumulated cost function. The discount factor also helps to motivate a choice for the length of the horizon $T$ \cite{Put:B94}. The optimal control sequence $\hat{\V{u}}_{1:T}$ can finally be obtained based on the \ac{rh} control\vspace{-.5mm} policy
\begin{align}\label{eq:policy2}
    \hat{\V{u}}_{1:T} = \hspace{.6mm}\argmin_{\V{u}_{1},\dots,\V{u}_{T} \in \mathcal{U}^{M_1}} \hspace{.6mm} C(\V{u}_{1:T}).
    \vspace{1mm}
\end{align}
Note that for $T\rmv=\rmv1$, the \ac{rh} control policy is equal to the myopic control policy in \eqref{eq:policy1}.

The so-called forward propagation defined by the accumulated cost function in \eqref{eq:accumulatedCost} naturally leads to a tree search, where a new level of the tree is introduced for every $t$. Thus, every possible sequence of control vectors $\V{u}_{1:t-1}$ up to time $t$, leads to $|\mathcal{U}|^{M_1}$ possible sequences of control vectors at time $t$. This implies that the computational complexity of the tree search scales as $\mathcal{O}\big(|\mathcal{U}|^{TM_1}\big)$, i.e., exponentially in $T$ and $M_1$.
To reduce the complexity of the search, a suboptimal control sequence can be obtained by determining, at each timestep $t$, the control sequence of each mobile anchor $j \rmv \in \rmv \{1,\dots,M_1\}$ individually, assuming $\V{u}_{k+t}^{(j')} \rmv=\rmv 0$ for all the other mobile anchors $j' \rmv \in \rmv \{1,\dots,M_1\} \backslash \{j\}$. In particular, there are $M_1$ parallel searches and for every search, every possible sequence of control vectors $\V{u}^{(j)}_{1:t-1}$ up to time $t$, leads to $|\mathcal{U}|$ possible sequences of control vectors at time $t$. The computational complexity of this approximate search thus only scales as $\mathcal{O}(M_1|\mathcal{U}|^{T})$. To further reduce computational complexity, a pruning strategy can be used. Here, after processing at every tree level $t\!\in\! \{1,\dots, T\}$ the number of possible sequences of control vectors in every parallel search is limited to $V$. In particular, we only keep the $V$ sequences with different final anchor positions, which yield the largest accumulated cost $C(\V{u}_{1:t})$. It can be easily verified that the computational complexity of the approximated search with the additional pruning strategy only scales as\vspace{0mm} $\mathcal{O}(V M_1 |\mathcal{U}| T)$.

\section{Results}
\label{sec:results}

Next, we will present a numerical case study to demonstrate that active planning can significantly improve the localization accuracy in a 3-D scenario at a reasonably computational cost\vspace{-1mm}.

\subsection{Simulation Setup and Example Realization}\label{sec:setup}
We consider a network that consists of one static anchor, one mobile anchor, and one mobile agent, i.e., $M\rmv=\rmv2$, $M_1\rmv=\rmv1$, and $N\rmv=\rmv1$. All initial positions are random and uniformly distributed on the \ac{roi}.
The mobile agent represents a pedestrian that moves in the X-Y plane and has a constant Z-coordinate equal to 0. Hence, the state of the agent at time $k$ can be represented by the vector $\V{x}^{(i)}_k = [x^{(i)}_{1,k}\iist x^{(i)}_{2,k}\iist x^{(i)}_{3,k}\iist \dot{x}^{(i)}_{1,k}\iist \dot{x}^{(i)}_{2,k}]^{\text{T}}$. At time $k=0$, the prior covariance of $\V{x}^{(i)}_0$ is set to $\V{\Sigma}^{(i)}_{0} \rmv=\rmv \text{diag}(10^2,10^2,50^2,0.1,0.1)$, and the prior mean $\V{\mu}^{(i)}_{0}$ is sampled from a Gaussian distribution that has the true agent state as the mean and $\V{\Sigma}^{(i)}_{0}$ as the covariance matrix. The state transition function in \eqref{eq:statetransitionf} is defined as follows. The substate $\big[x^{(i)}_{1,k}\iist x^{(i)}_{2,k}\iist \dot{x}^{(i)}_{1,k} \iist \dot{x}^{(i)}_{2,k}\big]^{\text{T}}$ follows a constant velocity motion model \cite{BarWilTia:B11} with driving noise variance $\sigma^2_{\nu}\rmv=\rmv 0.001$m$^2$/s$^4$ while the Z-coordinate $x^{(i)}_{3,k}$ remains constant.

The initial positions of the anchors are determined by drawing samples from the \ac{pdf} that is uniform on the 3-D \ac{roi} defined as $[-100\ist\text{m}, \ist 100\ist\text{m} ]  \times [-100\ist\text{m}, \ist 100\ist\text{m}] \times [-100\ist\text{m}, \ist 100\ist\text{m}]$. The mobile anchor moves in 3-D according to \eqref{eq:controlf}. The dictionary $\mathcal{U}$ consists of $|\mathcal{U}|=19$ possible control vectors that all have a constant norm $\|\V{u}\| = 2\ist \text{m/s}$. The control vectors are chosen such that their angles form a uniform grid of spherical coordinates. The duration of a discrete timestep is $T_{\rmv\Delta}=1\ist\text{s}$.
\begin{figure}[ht!]
    \centering
    \begin{subfigure}{.95\columnwidth}
    \centering
    \includegraphics[scale=0.5]{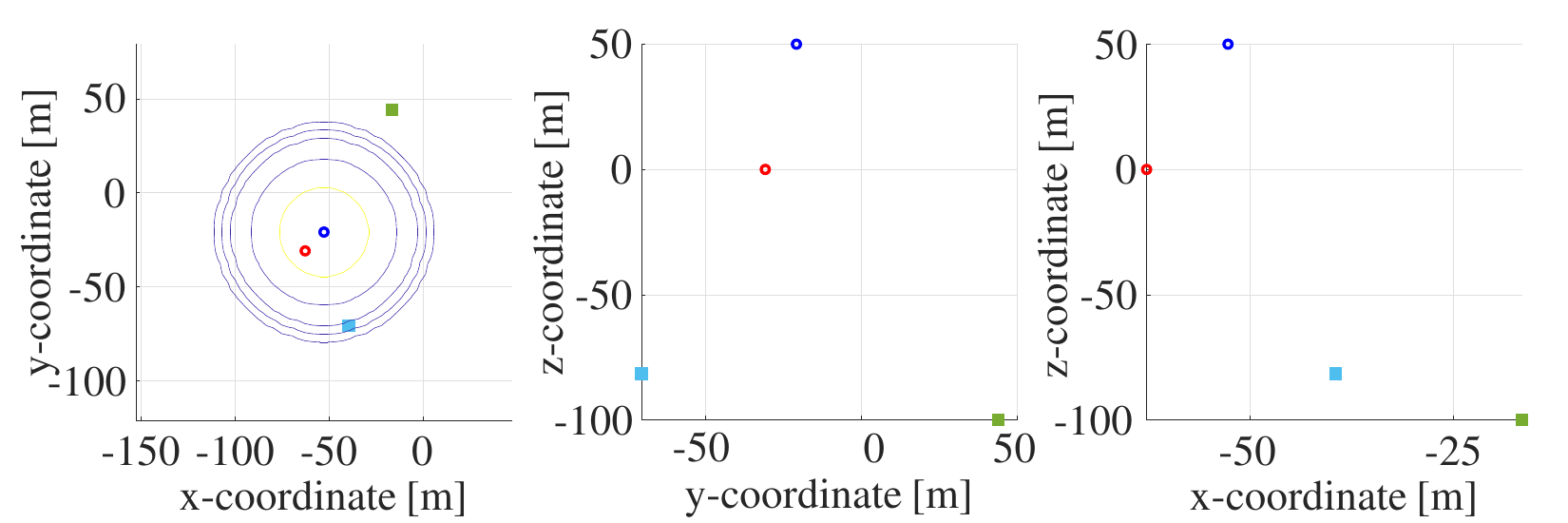}  
    \caption{Initial agent and anchor positions.}\label{subfig:init}
    \vspace{3mm}
    \end{subfigure}
    \hfill
    \begin{subfigure}{.95\columnwidth}
    \centering
    \includegraphics[scale=0.5]{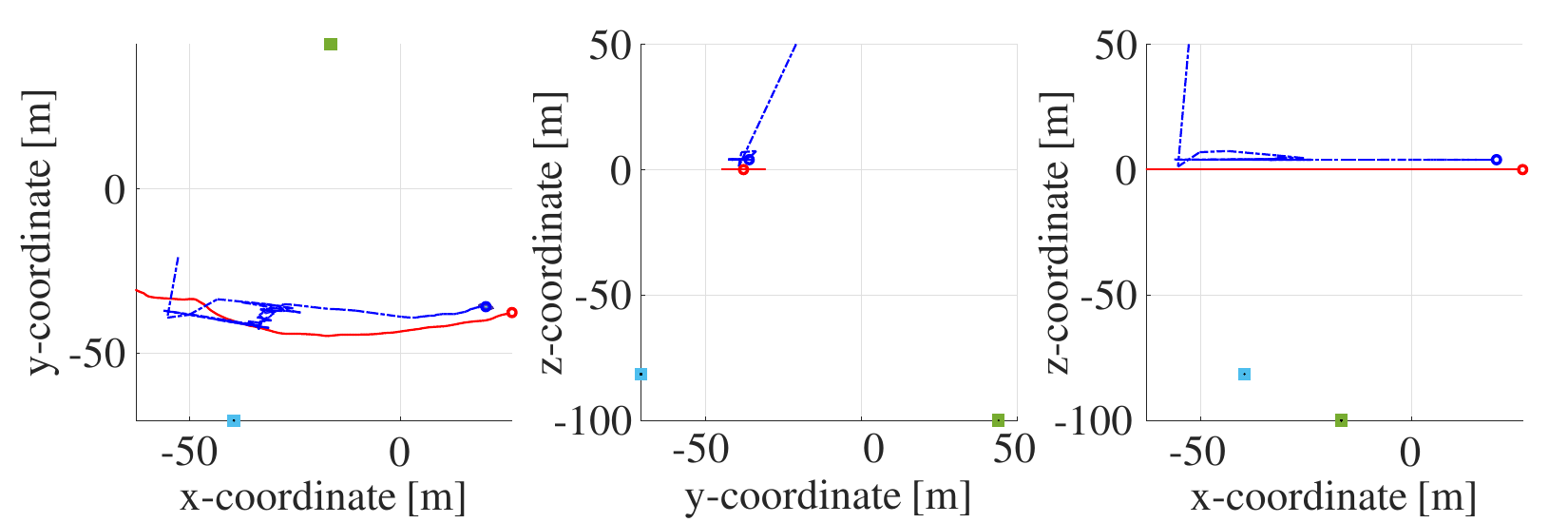}  
    \caption{Two static anchors: True and estimated agent track.}\label{subfig:static}
    \vspace{3mm}
    \end{subfigure}
    \hfill
    \begin{subfigure}{.95\columnwidth}
    \centering
    \includegraphics[scale=0.5]{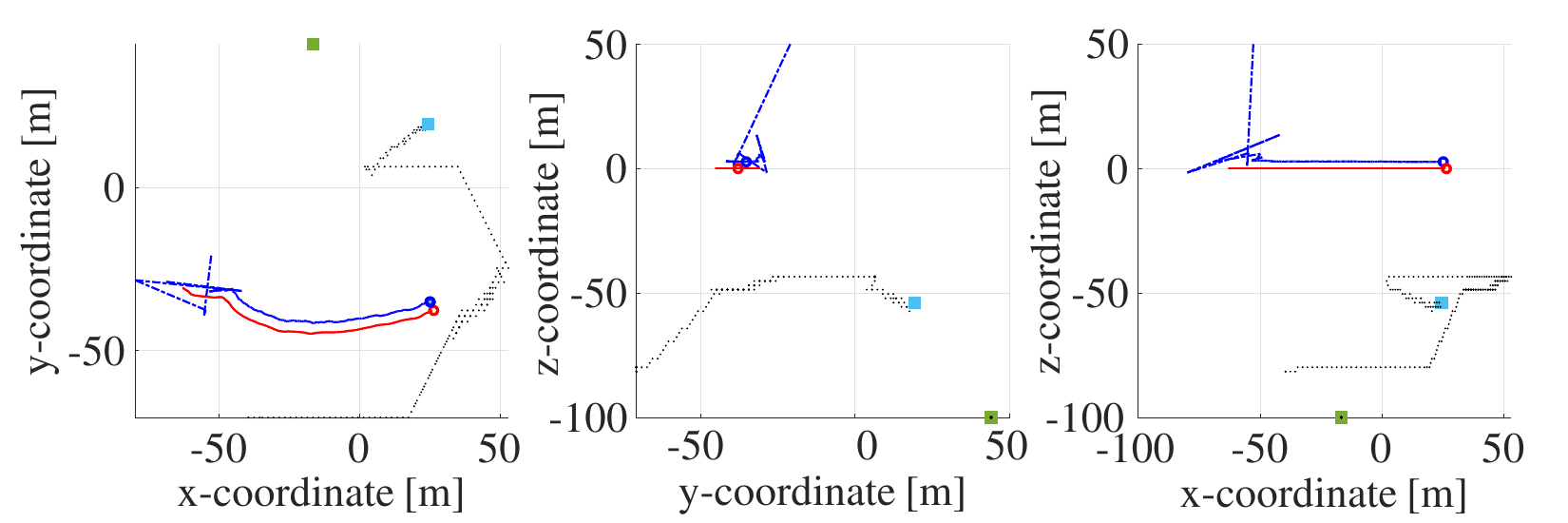}  
    \caption{Myopic anchor control: True and estimated agent track.}\label{subfig:myopic}
    \vspace{3mm}   
    \end{subfigure}
    \hfill
    \begin{subfigure}{.95\columnwidth}
    \centering
    \includegraphics[scale=0.5]{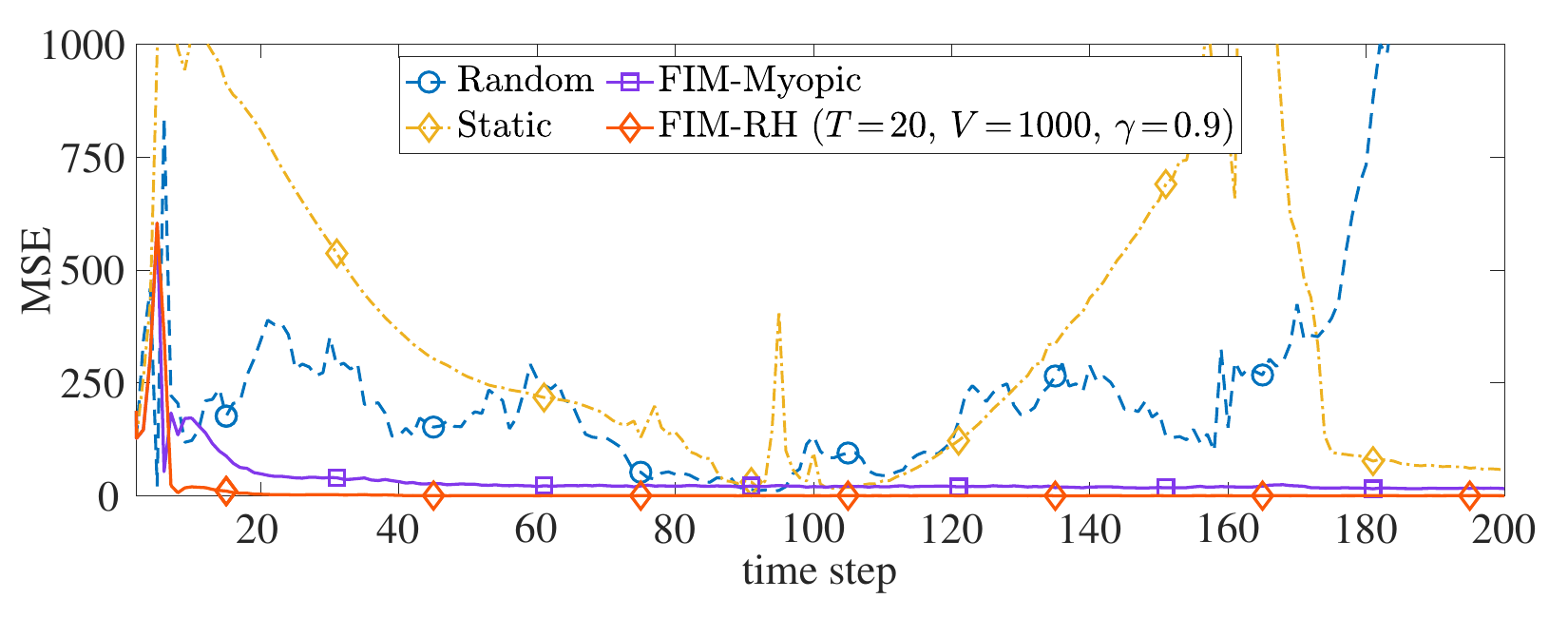}  
    \caption{MSE of the agent state estimate versus time.}\label{subfig:mse}
    \vspace{4mm}   
    \end{subfigure}
    \centering
    \caption{\small Example realization of the simulated 3-D localization scenario. The 3-D environment is mapped onto three planes. The two anchor positions are shown as green and blue squares, respectively. The red dot denotes the true position of the agent. The blue dot denotes the estimated position of the agent. The true and estimated agents tracks are shown as red and blue solid lines, respectively. The agent track is shown as a black dotted line. Error ellipses corresponding to the prior covariance matrix are also shown in the X-Y plane of (a). }\label{fig:demo}
    \vspace{-1mm}
\end{figure}
Range measurements are performed between each anchor and the agent following \eqref{eq:measurementf1}. The standard deviation of the measurement noise is $\sigma^{(1,1)}_{\omega} = \sigma^{(1,2)}_{\omega} = 0.1\ist \text{m}$. 

In addition to the proposed \ac{rh} control in \eqref{eq:policy2}, denoted ``FIM-RH'', we simulate the proposed myopic control in \eqref{eq:policy1}, denoted ``FIM-Myopic'', and two additional reference approaches. In ``static'', both anchors do not move. In ``random'', the mobile anchor moves randomly, i.e., at each timestep $k$, a control vector $\V{u}^{(1)}_{k+1}$ is randomly chosen from $\mathcal{U}$. For ``FIM-RH'', planning horizon is set as $T=20$, the pruning threshold to $V=1000$, and the discount factor to $\gamma=0.9$ For estimation, all methods rely on sigma point \ac{bp} \cite{MeyHliHla:J14}.

Fig.~\ref{fig:demo} shows an example realization that consists of 200 timesteps. Fig.~\ref{subfig:init} shows initial agent and anchor positions. In Fig.~\ref{subfig:static}, where ``static'' has been used, we can see that the localization accuracy is low and time-varying. In Fig. \ref{subfig:myopic}, FIM-Myopic'' has been used to control the mobile anchor. Here, we can see that the agent position estimate converges quickly to a position close to the true agent position and that the mobile agent can be tracked accurately. Fig. \ref{subfig:mse} shows the \ac{mse} of the agent state estimates versus time for this example realization that results from the different control methods. We can see that the proposed methods, i.e., ``FIM-Myopic'' and ``FIM-RH'', significantly outperform the two reference approaches. More specifically, we can see that the estimation accuracy is poor and time-varying for both the ``random'' and the ``static'' approach. On the contrary, the ``FIM-Myopic'' and ``FIM-RH'' methods, after a quick initial convergence, can accurately track the agents at all times. It can be noted that the ``FIM-RH'' method converges quicker and has a higher estimation accuracy compared to ``FIM-Myopic''. This is because ``FIM-RH'' can compute more effective control vectors by performing optimization over a receding planning\vspace{2mm} horizon.

\subsection{Numerical Results and Computational Complexity}
\begin{figure}[t]
    \centering
    \includegraphics[width=0.9\columnwidth]{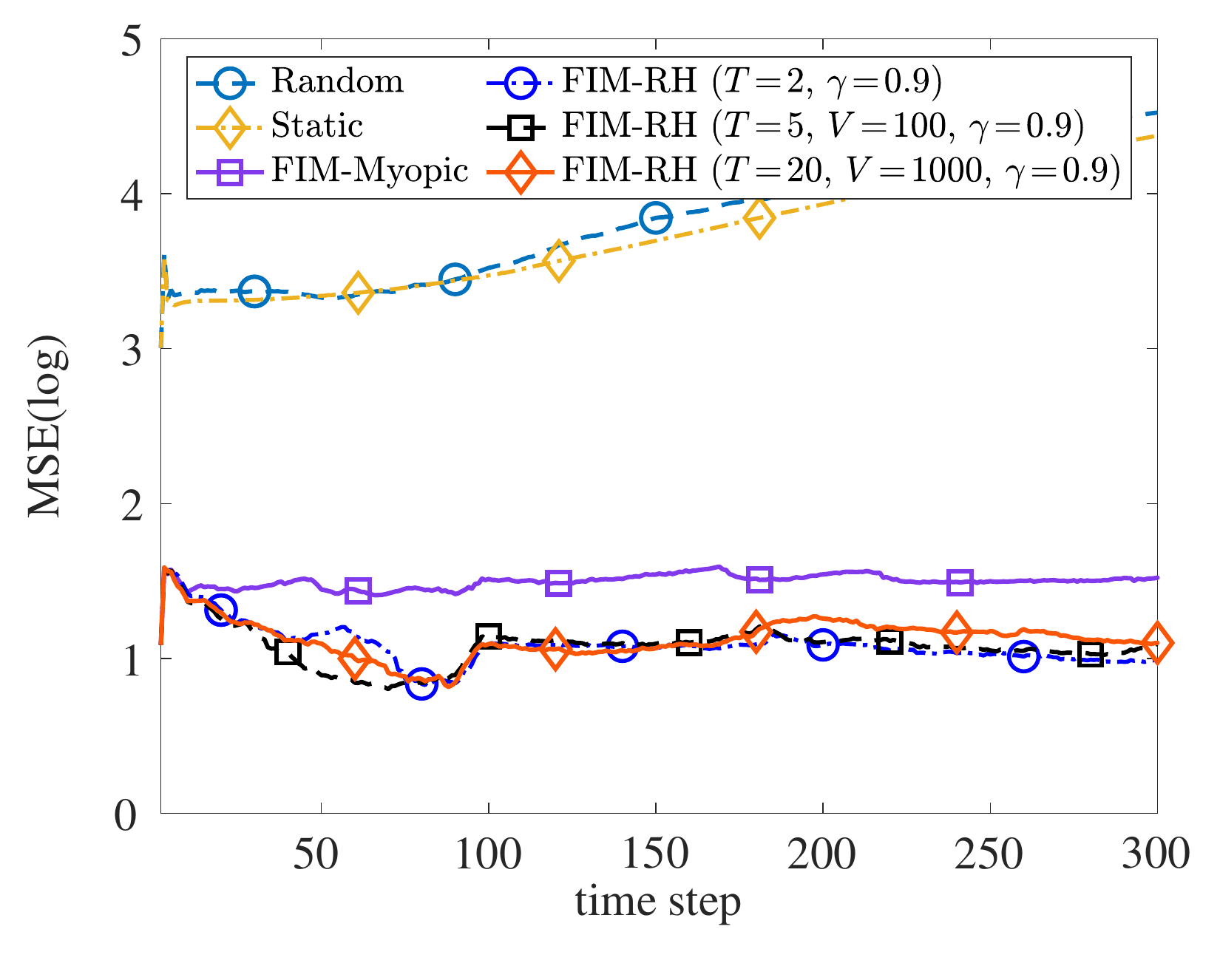}
    \centering
    \caption{Logarithmic \ac{mse} of agent position versus time. The proposed active planning methods result in a significant reduction of the \ac{mse}. }\label{fig:rmse1}
    \vspace{0mm}
\end{figure}

To statistically validate the localization accuracy, we performed 100 simulations that consist of 300 timesteps in the scenario discussed in Section \ref{sec:setup}. 
\subsubsection{Estimation Performance}
Fig.~\ref{fig:rmse1} shows the average base-10 logarithm of the \ac{mse} versus timestep for the different control strategies and the different \ac{rh} parameters. We can see that active planning significantly reduces the \ac{mse}. More specifically, neither the ``static'' nor the ``random'' approach works in this scenario, in the sense that the average logarithm of the \ac{mse} resulting from these two approaches is huge. On the contrary, ``FIM-Myopic'' and ``FIM-RH'' can accurately localize and track the agent. As expected, ``FIM-RH'' yields an improved accuracy compared to ``FIM-Myopic''.

To better understand the influence of the different \ac{rh} parameters in this scenario, we compare three different \ac{rh} parameter sets, i.e., $T=2$; $T=5, V=100$; and $T=20, V=1000$. The resulting estimation performance curves are shown in Fig.~\ref{fig:rmse1}. These results indicate that in the considered scenario, high localization accuracy can already be achieved for $T=2$, while $T=5, V=100$ and $T=20, V=1000$ perform very similarly. In this rather simple scenario, a planning horizon of $T=2$ is enough to leverage the full estimation performance, and thus increasing $T$ does not lead to improved accuracy. Longer planning horizons $T$ are expected to help in larger scenarios with more anchors and more agents.

\subsubsection{Lost Tracks} In the simulated scenario, we sometimes observe lost tracks, i.e., the agent position estimate deviates from the true agent position and thus the \ac{mse} grows with time. These track losses are related to the fact that we initialize the agent and anchor positions randomly, which can lead to geometries that are not favorable for estimating the position of the agent. This results in a large \ac{mse} at initial time steps and the possibility that the estimation error grows with time. We declare a track loss if the \ac{mse} exceeds a value of 1000 at any timestep between timestep 3 to 50.

To determine the track loss percentage, we simulated 10000 random trails with 5 different methods, i.e., ``static'', ``random'', ``FIM-Myopic'', ``FIM-RH ($T\rmv=2,\gamma\rmv=\rmv0.9$)'' and ``FIM-RH ($T\rmv=\rmv5,V\rmv=\rmv100,\gamma\rmv=\rmv0.9$)''. Note that we have not simulated ``FIM-RH ($T\rmv=\rmv20,V\rmv=\rmv1000,\gamma\rmv=\rmv0.9$)'' since this would result in an excessive runtime as discussed below. Table \ref{tab:trackloss} shows the track loss ratio of all methods. While the reference methods ``Static'' and ``Random'' result in more than 40\% of lost tracks, the proposed methods can reduce the lost tracks to less than 1\%. This demonstrates that the use of active planning can counteract initial network topologies that are unfavorable for agent\vspace{.5mm} localization
\begin{table}[ht]
    \centering
    \begin{tabular}{ |p{5cm}||p{2.5cm}|  }
    \hline
    Setting     & Lost Tracks (\%)\\
    \hline
    Static  & 44.24 \\
    Random  & 46.02 \\
    FIM-Myopic  & 13.72 \\
    FIM-RH ($T=2,\gamma=0.9$)   & 0.66 \\
    FIM-RH ($T=5,V=100,\gamma=0.9$)   & 0.62 \\
    \hline
    \end{tabular}
    \vspace{0mm}
\caption{Percentage of lost tracks for all simulated methods.}\label{tab:trackloss}
\end{table}

\subsubsection{Computational Complexity}
The simulation is implemented using MATLAB on a 2017 MacBook Pro with an Intel Core-i5 processor. The runtime for a single run in the considered simulation scenario with 200 timesteps is 1.1 seconds for ``FIM-Myopic'', 4.8 seconds for ``FIM-RH ($T=2, \gamma=0.9$)'', 23 seconds for ``FIM-RH ($T=5, V=100, \gamma=0.9$)'', and 1964 seconds for ``FIM-RH ($T=20,V=1000,\gamma=0.9$)''. This demonstrates the low computational complexity of our method for short planning horizons $T$. However, as also analyzed in Section \ref{sec:RH}, the complexity increases rapidly with increasing $T$, which makes the current approach unsuitable for large networks.
\pagebreak

\section{Conclusion}
\label{sec:conclusion}
In this paper, we presented an active planning method that controls mobile anchors in a localization network with the goal to maximize information gain of future measurements. The traces of approximate inverse Bayesian \acp{fim} were used to develop an efficient cost function for active planning that is suitable for \ac{rh} control. Minimizing the cost function can be formulated as a tree-search problem. We also discuss simple approximation strategies that make it possible to solve the resulting tree-search problem efficiently. A numerical case study demonstrates that active planning based on \ac{rh} control can lead to performance improvements compared to a myopic control. The main limitation of the current method is that the approximation of the Bayesian \ac{fim} does not take the measurements of the other agents into account \cite{SheWymWin:10}. In addition, the current search strategies are expected to be highly suboptimal in large networks. Future work will address these limitations. In particular, we expect that advanced learning methods can lead to a near-optimal \ac{rh} control in large networks. We also plan to develop active planning for estimation problems with data association uncertainty \cite{MeyKroWilLauHlaBraWin:J18,MeyWil:J21}.

\section*{Acknowledgement}
DISTRIBUTION STATEMENT A: Approved for public release. This material is based upon work supported by the Under Secretary of Defense for Research and Engineering under Air Force Contract No. FA8702-15-D-0001. Any opinions, findings, conclusions, or recommendations expressed in this material are those of the author(s) and do not necessarily reflect the views of the Under Secretary of Defense for Research and \vspace{1mm} Engineering.

\renewcommand{\baselinestretch}{1.05}
\selectfont
\bibliographystyle{IEEEtran}
\bibliography{StringDefinitions,IEEEabrv,Papers,Books,Temp}

\end{document}